\documentclass{article}

     \usepackage[preprint]{neurips_2020}

\usepackage[utf8]{inputenc} 
\usepackage[T1]{fontenc}    
\usepackage{hyperref}       
\usepackage{url}            
\usepackage{booktabs}       
\usepackage{amsfonts}       
\usepackage{nicefrac}       
\usepackage{microtype}      
\usepackage[pdftex]{graphicx}
\usepackage{subfigure}
\usepackage{amsmath}

\setcitestyle{numbers}

\usepackage{color,soul}

\title{Machine Learning-driven Autotuning of Graphics Processing Unit Accelerated Computational Fluid Dynamics for Enhanced Performance}

\author{%
  Weicheng Xue \\
  Kevin T. Crofton Department of Aerospace and Ocean Engineering\\
  Virginia Tech\\
  Blacksburg, VA, 24060\\
  \texttt{weich97@vt.edu} \\
   \And
  Christopher J. Roy\\
  Kevin T. Crofton Department of Aerospace and Ocean Engineering\\
  Virginia Tech\\
  Blacksburg, VA, 24060\\
  \texttt{cjroy@vt.edu}\\
}

\begin{document}

\maketitle

\begin{abstract}

Optimizing the performance of computational fluid dynamics (CFD) applications accelerated by graphics processing units (GPUs) is crucial for efficient simulations. In this study, we employed a machine learning-based autotuning technique to optimize 14 key parameters related to GPU kernel scheduling, including the number of thread blocks and threads within a block. Our approach utilizes fully connected neural networks as the underlying machine learning model, with the tuning parameters as inputs to the neural networks and the actual execution time of a simulation as the outputs. To assess the effectiveness of our autotuning approach, we conducted experiments on three different types of GPUs, with computational speeds ranging from low to high. We performed independent training for each GPU model and also explored combined training across multiple GPU models. By leveraging artificial neural networks, our autotuning technique achieved remarkable results in tuning a wide range of parameters, leading to enhanced performance for a CFD code. Importantly, our approach demonstrated its efficacy while requiring only a small fraction of samples from the large parameter search space. This efficiency is attributed to the effectiveness of the fully connected neural networks in capturing the complex relationships between the parameter settings and the resulting performance. Overall, our study showcases the potential of machine learning, specifically fully connected neural networks, in autotuning GPU-accelerated CFD codes. By leveraging this approach, researchers and practitioners can achieve high performance in scientific simulations with optimized parameter configurations.
\end{abstract}

\section{Introduction}

Graphics processing units (GPUs), as highlighted in the study by Hwu et al.\cite{hwu2011gpu}, have garnered significant attention in the field of scientific computing due to their enhanced computing capabilities and higher memory throughput in comparison to central processing units (CPUs). CPUs typically serve as hosts that handle general settings and controls, while GPUs act as accelerator devices that execute intensive computations to achieve speedups. The GPU's architecture, featuring thousands of lightweight cores, enables faster and more parallel computation synchronously on the device. Once the device completes the computations, the results are transferred back to the host. Consequently, data movements occur between the host and the device due to their distinct memories. The host and device can be interconnected through PCI-E or NVLink\cite{li2019evaluating}, which evidently enhances the memory bandwidth.

\begin{figure}[ht]
  \centering
  \includegraphics[width=.6\linewidth,  trim = 7 7 7 7, clip]{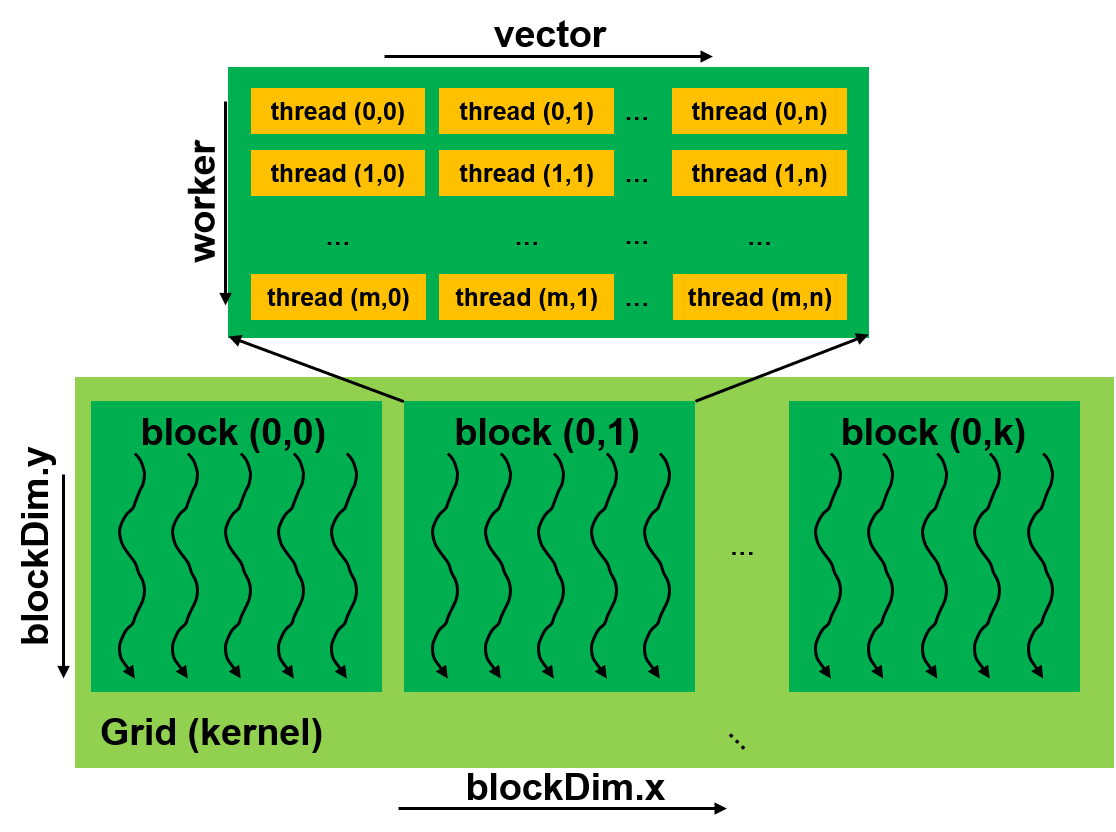}
  \caption{Multilevel parallelism for GPU.}
  \label{multi-para}
\end{figure}

GPU exhibits multiple levels of parallelism, including block-level and thread-level parallelism, the tuning parameters of which are the block size $k$, the worker size $m$, and the vector length $n$ in this example, as illustrated in Fig.~\ref{multi-para}. In GPU programming, the execution of a kernel function is orchestrated by a grid, which comprises a set of blocks or gangs. Each block consists of a group of threads that execute concurrently on the GPU's streaming multiprocessors (SMs). Within this parallel execution model, a worker represents the smallest unit of execution on the GPU, typically referring to individual threads or processing elements responsible for carrying out specific computations. Threads within a warp, a fundamental unit of execution in GPU architecture, are managed and scheduled together by the GPU's hardware, enabling efficient execution of instructions. Collectively, these threads, also referred to as vectors, represent individual instances of execution within the GPU, leveraging the device's parallel processing capabilities to accelerate various computational tasks. For these parallel structures, optimization of parameters such as blockDim.x, blockDim.y, worker size, and vector length is essential to achieve optimal performance. It is important to note that various GPU parallel standards such as CUDA~\cite{CUDA}, OpenCL~\cite{OpenCL}, OpenACC~\cite{OpenACC}, or OpenMP offload~\cite{mishra2017benchmarking} may employ different terminology for scheduling structures. However, the underlying concepts remain similar, and these scheduling structures can be mapped to the same hierarchical layers. To achieve high performance across different platforms with varying architectures or in different environments involving diverse problem sizes, numerical schemes, etc., users need to frequently tune their GPU-accelerated code to target devices. A single program may comprise more than 10 or even 100 kernels, each with multiple tuning parameters. Each parameter can have a wide range of possible values, resulting in an extremely large search space. It can be challenging for domain scientists with limited experience to effectively tune these parameters. Additionally, while a compiler may offer default settings to assist with parameter tuning, the default configuration is unlikely to be the optimal choice.

In this work, a computational fluid dynamics (CFD) code is accelerated on various types of GPU using OpenACC. OpenACC is a library specification specifically designed for GPU programming. Originally, it was developed as an extension of OpenMP to support GPUs. However, due to the delayed full implementation of OpenMP 4.0~\cite{sultana2016openacc} or higher, the development of OpenACC has become increasingly independent. OpenACC serves as a valuable tool to simplify programming on heterogeneous CPU/GPU computing systems, which is a significant factor behind its utilization in this work.

When employing OpenACC for optimizing a CFD code, it becomes imperative to fine-tune various parameters, such as gang size and vector length, to achieve optimal performance on GPU architectures. In the context of OpenACC, the term "gang" typically refers to a group of threads, and "vector length" corresponds to the number of threads within a gang. These parameters are akin to CUDA thread block size and CUDA thread number, respectively.

The optimization process involves a delicate balance. On one hand, it is crucial to harness a sufficient number of threads and gangs to effectively utilize the GPU, thereby maximizing utilization and concurrency. On the other hand, the GPU's streaming multiprocessor registers are shared among all threads within a block, and an excessively high number of concurrently running threads may adversely impact concurrency.

\begin{figure}[ht]
  \centering
  \includegraphics[width=.6\linewidth,  trim = 7 7 7 7, clip]{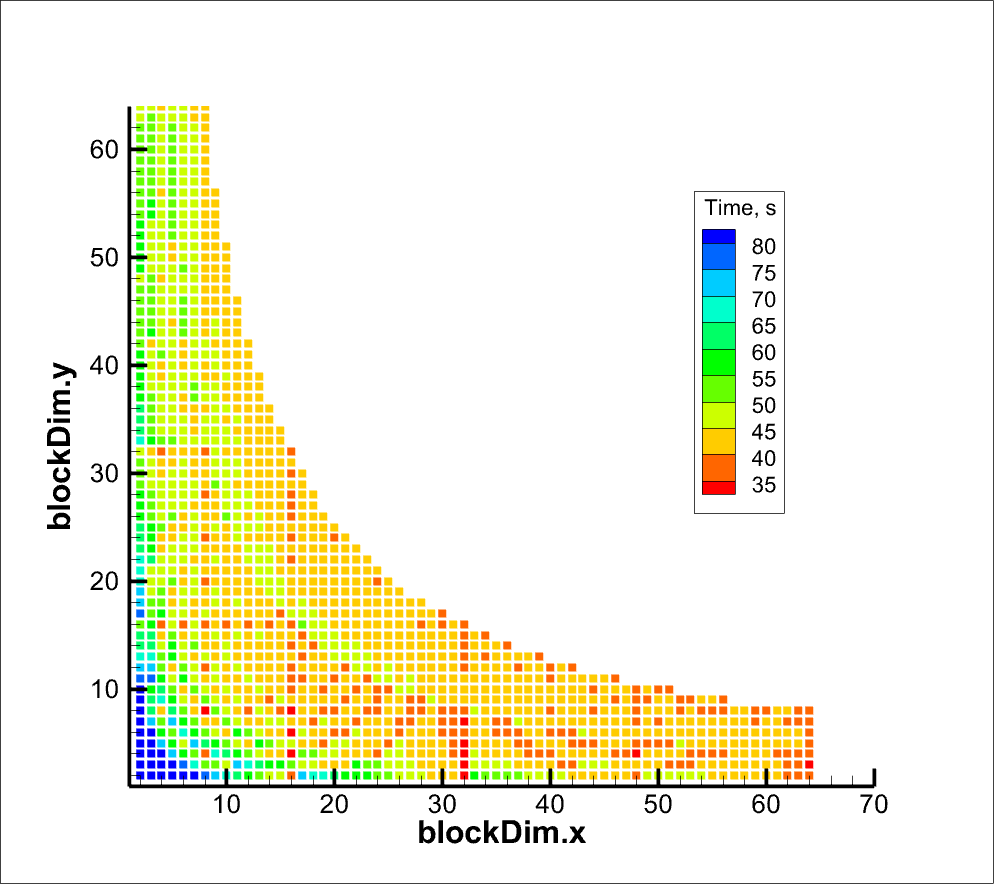}
  \caption{Two-parameter manual tuning for a buoyancy-driven cavity code.}
  \label{ldc-tuning}
\end{figure}

\begin{figure}[ht]
	\centering 
	\subfigure[3D Buoyancy driven cavity solution]{ 
		\label{3D_Pressure}
		\includegraphics[width=.45\textwidth,trim=0 0 0 0,clip]{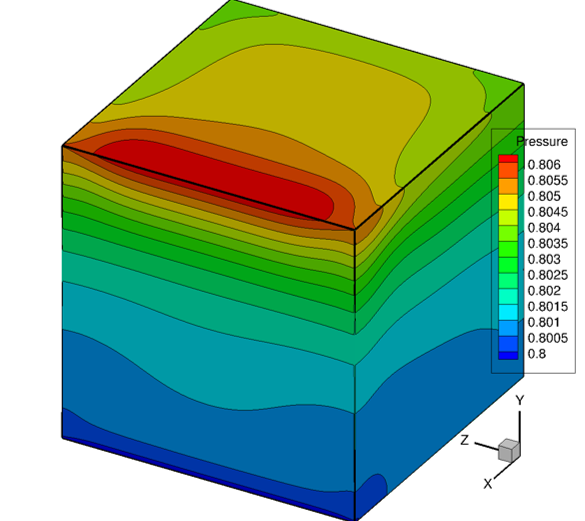} 
	} 
	\subfigure[2D Buoyancy driven cavity solution]{ 
		\label{2D_Temperature}
		\includegraphics[width=.45\textwidth,trim=0 0 0 0,clip]{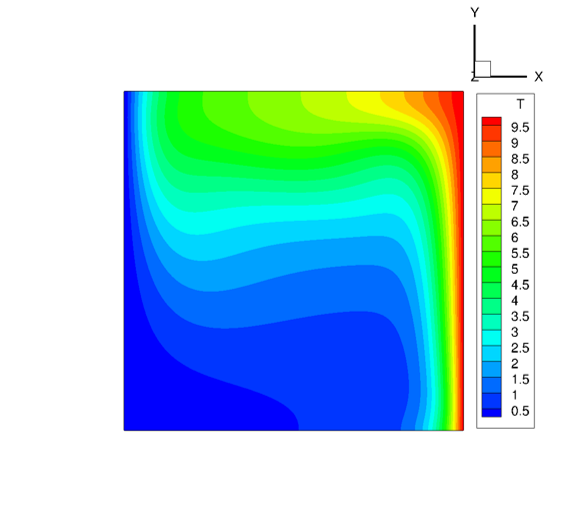} 
	} 
	\caption{Buoyancy driven cavity cases} 
	\label{BDC}
\end{figure}

The illustration in Fig.~\ref{ldc-tuning} depicts a two-parameter manual tuning of the thread-block size for a buoyancy-driven cavity code. With only two tuning parameters, blockDim.x and blockDim.y, there are over 200 possible configurations to consider. Quickly identifying a fairly optimal configuration through manual tuning can be challenging, especially in cases involving multiple kernels and additional tuning parameters. The sheer number of combinations underscores the complexity of the tuning process and highlights the need for automated or systematic approaches to efficiently navigate through the optimization space. Fig.~\ref{BDC} shows the 3D pressure contour solution and the 2D temperature contour solution for the buoyancy-driven problem on a $64 \times 64 \times 64$ grid and a $64 \times 64$ grid, respectively. The 3D/2D buoyancy driven cavity problem has a 3D/2D cubic domain, a vertical wall and its opposing wall have different temperatures, and the horizontal walls are adiabatic. A gravitational force is added to the air in the square cavity. Heat flux caused by the temperature difference leads to small density changes in the fluid (Boussinesq approximation), and the buoyancy effect (density change) causes the fluid to convect in the cavity. The buoyancy-driven cavity problem is solved using an artificial viscosity method developed by Chorin~\cite{chorin1997numerical}. More details about the numerical schemes and boundary conditions can be found in Ref~\cite{xue2021multi}. While compilers can automatically determine how to map loop iterations to different levels of parallelism on the device, a well-designed autotuning approach can further improve scheduling. Autotuning becomes particularly crucial for achieving peak performance across diverse domains on next-generation GPUs.

Furthermore, the optimal tuning strategy may vary based on the characteristics of the application. For example, a high-latency application may necessitate a higher occupation to increase concurrent accesses and mask latency, whereas a low-latency application could benefit from higher concurrency to exploit more computational capabilities. Notably, the compiler is likely to choose similar settings for distinct problems, as factors like problem size are runtime variables that cannot be determined at compile time. Thus, manual or automated tuning is essential for adapting to the specific requirements of different computational workloads.

\begin{figure}[ht]
  \centering
  \includegraphics[width=1.0\linewidth,  trim = 7 0 0 7, clip]{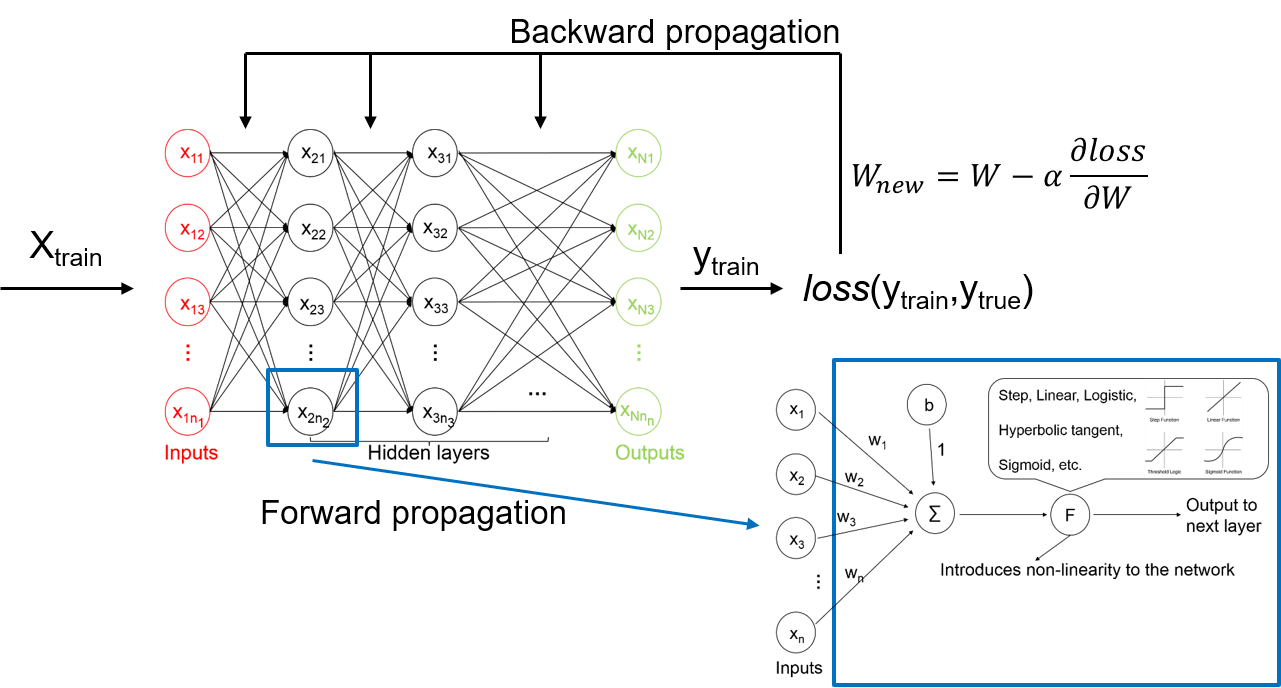}
  \caption{Artificial Neural Network}
  \label{ANN}
\end{figure}

Artificial neural networks have shown great promise as methods for autotuning. A typical artificial neural network consists of multiple layers of neurons, where each neuron performs a set of arithmetic operations. These operations typically involve calculating the dot product between the weights and inputs. Activation functions are used to introduce non-linearity to each layer. The output from each layer is then propagated to the next layer, a process known as forward propagation. During training, the loss function is calculated by comparing the predicted data with the ground truth. To improve the generalization of the trained model, a penalty function is often appended to the loss function. The weights for each layer can be updated using the chain rule and the jacobians of the loss function with respect to the weights. This process, known as backward propagation, involves using different types of optimizer schemes to update the weights. The entire procedure is depicted in Fig.~\ref{ANN}, where the forward propagation and backward propagation steps are illustrated.
 
In this work, a neural network is employed to estimate the runtime of the GPU-accelerated CFD solver. Tuning parameters, including gang sizes and vector lengths for different kernels, are used as input features for the neural network, while the CFD solver runtime serves as the target output. Given that the execution time of the program is continuous-valued, this problem falls into the category of supervised regression learning. To generate the training and testing datasets, a moderate number of samples (7500 samples for each GPU model) are generated by executing the CFD code. The fully connected neural network utilized in this research consists of multiple hidden layers following the input layer. Each hidden layer is composed of a number of neurons activated by the rectified linear unit (ReLU) function. Hyperparameters such as the number of layers, number of neurons in each layer, learning rate, etc., are tuned to obtain a well-trained model. Finally, the trained model is evaluated on the testing dataset to demonstrate the effectiveness of the machine learning-based autotuning approach.   

\section{Related Work}

There are various approaches for autotuning code to obtain better performance. Pickering et al.~\cite{pickering2015directive} performed an exhaustive search of the entire space of 2D thread-block dimensions for a GPU-accelerated test case to identify the optimal configuration. They also discovered that the compiler default cannot guarantee optimal configurations in different scenarios. However, manual tuning requires significant expertise, time, and is prone to errors. In addition, their work is only focused on tuning the GPU thread-block size of a single kernel, which consists of only two parameters. Jia et al.~\cite{jia2013starchart} employed a statistic tree-based approach to develop a tool called Starchart, to adaptively explore the parameter space, identifying the most influential tuning parameters and their ranges. In addition, by employing recursive partitioning regression trees, their approach can automatically determine the optimal hardware and software configurations for a given application. However, the effectiveness of Starchart heavily relies on the availability and quality of training data. Also, as the complexity and dimensionality of the parameter space increase, the computational cost and time required for training and searching in Starchart may become prohibitively high. Collins et al.~\cite{collins2013masif} utilized a machine learning-based method integrated with principal components analysis to significantly reduce the search space on parallel skeletons, which are high-level abstractions of parallel computation patterns. However, if the models are overly specialized to the training data, MaSiF can lead to poor generalization and suboptimal performance on unseen data. Falch et al.~\cite{falch2017machine} employed neural networks for autotuning OpenCL kernel applications and achieved good predictions for certain benchmark codes. However, their autotuning framework exhibited limited robustness, as the optimal configuration could vary greatly across multiple runs. This indicates that the autotuning framework might not guarantee a robust and globally optimal solution for all scenarios. Cui et al.~\cite{cui2020iterml} developed an iterative machine learning approach that seeks potentially superior samples in subsequent iterations based on samples from previous iterations. They did experiments on the GPU thread-block size across many benchmarks running on an NVIDIA P100 or V100 GPU and showed that their automated iterative machine learning approach can reduce search effort by 40\% to 80\% when compared to non-iterative ML. Chen et al.~\cite{chen2018tvm} presented a learning-based framework called TVM to optimize tensor programs for deep learning workloads. TVM uses machine learning techniques to efficiently explore a large space of tuning configurations and automatically selects the best configurations for optimal performance. Schoonhoven et al.~\cite{schoonhoven2022benchmarking} conducted a survey by performing experiments on a large number of kernel spaces, GPUs and evolutionary black-box optimization algorithms. They introduced a metric based on the PageRank centrality for gaining insight into the difficulty of the optimization problem and demonstrated the novel metric can be used to represent the observed tuning performance accurately.

\section{Data Collection}

\subsection{The CFD Code Base: SENSEI}

SENSEI (Structured, Euler/Navier-Stokes Explicit-Implicit Solver) is a modern CFD code developed in modern Fortran. This code employs a multi-block finite volume method to solve the governing equations, which can be expressed in the weak form as shown in Eq.~(\ref{governing}).
 
\begin{equation}
\label{governing}
\frac{\partial }{\partial t}\int_\Omega \vec{Q}{\rm d}\Omega +\oint_{\partial \Omega} (\vec{F_{i,n}}-\vec{F_{\nu,n}}){\rm d}A= \int_\Omega \vec{S}{\rm d}\Omega
\end{equation}
where $\vec{Q}$ represents the vector of conserved variables, and $\vec{F_{i,n}}$ and $\vec{F_{\nu,n}}$ are the inviscid and viscous flux normal components (the dot product of the 2nd order flux tensor and the unit face normal vector), respectively. The source term $\vec{S}$ incorporates contributions from body forces, chemistry source terms, or the method of manufactured solutions~\cite{oberkampf2010verification}. The definition of $\vec{Q}$, $\vec{F_{i,n}}$ and $\vec{F_{\nu,n}}$ are given in Eq.~(\ref{QF})
\begin{equation}
\label{QF}
\vec{Q}=
\begin{bmatrix}
\rho\\\rho u\\\rho v\\\rho w\\\rho e_t
\end{bmatrix}, \,
\vec{F_{i,n}}=
\begin{bmatrix}
\rho V_n\\\rho u V_n + n_x p\\\rho v V_n + n_y p\\\rho w V_n + n_z p\\\rho h_t V_n
\end{bmatrix}, \,
\vec{F_{\nu,n}}=
\begin{bmatrix}
0\\n_x \tau_{xx} + n_y \tau_{xy} + n_z \tau_{xz}\\n_x \tau_{yx} + n_y \tau_{yy} + n_z \tau_{yz}\\n_x \tau_{zx} + n_y \tau_{zy} + n_z \tau_{zz}\\
n_x \Theta_{x} + n_y \Theta_{y} + n_z \Theta_{z}
\end{bmatrix}	    
\end{equation}
where $\rho$ is the density, $u$, $v$, $w$ are the Cartesian velocity components, $e_t$ is the total energy, $h_t$ is the total enthalpy, $V_n = n_x u+ n_y v + n_z w$ and the $n_i$ terms are the components of the outward-facing unit normal vector of the domain boundary $\partial \Omega$. $\tau_{ij}$ are the viscous stress components based on Stokes's hypothesis. $\Theta_i$ represents the heat conduction and work from the viscous stresses.

The spatial computational domain $\Omega$ is discretized into finite control volumes $\Omega_{k}$, forming the union of all control volumes. The weak form in Equation (\ref{governing}) is redefined for each finite control volume in Equation (\ref{governing_cv}).

\begin{equation}
\frac{\partial }{\partial t}\int_{\Omega_{k}} \vec{U}{\rm d}\Omega +\oint_{\partial \Omega_{k}} (\vec{F_{i,n}}-\vec{F_{\nu,n}}){\rm d}s= \int_{\Omega_{k}} \vec{S}{\rm d}\Omega.
\label{governing_cv}
\end{equation}

The discrete solution, denoted as $\Vec{U}_{h}$, assumes constancy within each control volume. The semi-discrete form of the conservation law, expressed in Equation (\ref{spatial_residual}), involves the volume $|\Omega_{k}|$, the cell averaged solution vector $\Vec{U}_{h}$, and the spatial residual vector $\Vec{R}_{h}$.

\begin{equation}
  |\Omega_{k}|\dfrac{\partial}{\partial t}\vec{U}_{h} + \Vec{R}_{h} = \vec{0},
  \label{spatial_residual}
\end{equation}

The spatial residual, as defined in Equation (\ref{residual}), involves cell face contributions and the cell-averaged source term vector. The discretized equations can be advanced in time using an ODE solver, provided stability conditions are met.

\begin{equation}
  \Vec{R}_{h} = \sum_1^{f} (\vec{F_{i,n}}-\vec{F_{\nu,n}}){\rm \Delta}s - |\Omega_{k}| \vec{S_h}, 
  \label{residual}
\end{equation}
where $f$ is the cell face number ($f = 4$ and $f = 6$ for 2D and 3D, respectively), $\Delta s$ is the face area, and $S_{h}$ is the cell averaged source term vector.

\begin{figure}[ht]
	\centering
	\includegraphics[width=.8\textwidth]{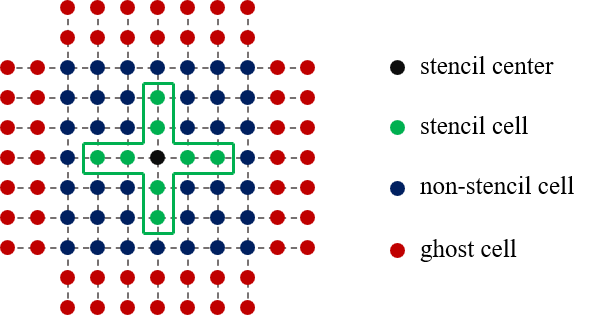}
	\caption{Stencil for MUSCL extrapolation}
	\label{stencil}
\end{figure}

The calculation of residuals centers around reconstructing fluxes using MUSCL extrapolation, detailed in Equation (\ref{flux}). MUSCL extrapolation extends conserved variables to cell faces within a stencil, as depicted in Fig.~\ref{stencil}.

\begin{equation}\label{flux}
\begin{aligned}
\vec{Q}_{k+1/2}^L &= \vec{Q}_k + \frac{\epsilon}{4} [(1 - \kappa) \Psi_{k-1/2}^{+} (\vec{Q}_{k} - \vec{Q}_{k-1}) + (1 + \kappa) \Psi_{k+1/2}^{-} (\vec{Q}_{k+1} - \vec{Q}_{k})] \\
\vec{Q}_{k+1/2}^R &= \vec{Q}_{k+1} - \frac{\epsilon}{4} [(1 + \kappa) \Psi_{k+1/2}^{+} (\vec{Q}_{k+1} - \vec{Q}_{k}) + (1 - \kappa) \Psi_{k+3/2}^{-} (\vec{Q}_{k+2} - \vec{Q}_{k+1})]
\end{aligned}
\end{equation}

where $\epsilon$ and $\kappa$ are MUSCL extrapolation parameters, $\Psi$ are limiter function values. $L$ and $R$ denote the left and right states, respectively. The accuracy of the flux is regulated by the variable $\epsilon$ in the computational scheme, with this work employing $\epsilon = 1$ for a second-order accuracy in space. Additionally, the variable $\kappa$ plays a crucial role in determining the nature of the scheme: whether it is fully upwind, upwind-biased, third-order, Leonard's Quick scheme, or central difference. Notably, the choice of a fully upwind scheme is common in scenarios featuring discontinuities.

SENSEI is versatile, capable of solving Euler equations, laminar flows, and Favre-Averaged Navier-Stokes equations with various turbulence models, incluiding Spalart-Allmaras\cite{allmaras2012modifications} and Menter's Shear stress transport models~\cite{menter1994two,menter2003ten}. It also incorporates functionalities to estimate truncation errors for high-order discretization error calculations~\cite{wang2020error,wang2022iterated}.

To enhance performance, SENSEI utilizes struct-of-array (SOA) memory layout instead of array-of-struct (AOS), promoting coalesced memory access. It is GPU-accelerated, achieving significant speedup compared to CPU counterparts. Random sampling is employed to efficiently autotune SENSEI's performance using machine learning, addressing the challenges posed by numerous kernels and tuning parameters.

SENSEI exhibits remarkable performance gains through parallelization on multiple GPUs, achieving an impressive speedup of up to 90$\times$ when employing 16 V100 GPUs in comparison to 16 Xeon CPU E5-2680v4 cores~\cite{xue2021improved}. Moreover, the code benefits from CPU-GPU heterogeneous computing, outperforming pure GPU computing with a notable speedup exceeding 20\%\cite{xue2023cpu}. Despite these achievements, SENSEI's performance is yet to be fully optimized, mainly due to the intricacies associated with numerous kernels, each featuring multiple tuning parameters such as gangs and vectors. These parameters can serve as input features, contributing to an expansive search space. The complexity is further compounded by potential inter-dependencies among discrete parameters\cite{feng2018deep}.

In our current endeavor, we employ a systematic random sampling methodology to create a diverse set of samples that comprehensively cover the expansive search space. These samples are judiciously partitioned into a training set, constituting 75\% of the data, and a testing set comprising the remaining 25\% of unseen samples. This strategic approach is geared towards effectively addressing the challenge of autotuning SENSEI's performance. By harnessing the capabilities of machine learning, we aim to efficiently optimize the code's performance, providing a robust and adaptable performance gain across various generations of GPUs..

\subsection{A Test Case}

Through meticulous examination of various 2D cases, we have determined that the foundational workflow remains consistently applicable regardless of the specific case, albeit with unique optimal configurations. While different numerical schemes may manifest variations in computational hotspots, users are initially tasked with executing a specific case for a brief period and identifying key kernels that significantly impact processing speed to enhance the efficacy of auto-tuning. Subsequently, the final step entails auto-tuning the parameters for these kernels. This process underscores the independence of auto-tuning from specific cases. Thus, although demonstrating similar outcomes and drawing analogous conclusions regarding the utility of machine learning (ML) in accelerating GPU-accelerated code may seem repetitive, it plays a pivotal role in affirming the robustness of our methodology.

\begin{figure}[ht]
	\centering
	\includegraphics[width=.6\textwidth,trim=7 7 7 7,clip]{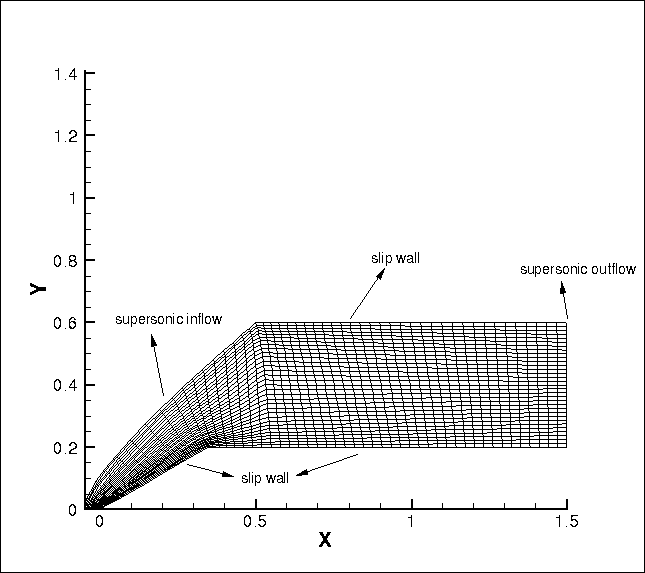}
	\caption{A coarse ($52 \times 16$) grid for the 2D inlet flow}
	\label{inlet-grid}
\end{figure}

In this study, we use a specific test case known as a 2D 30-degree supersonic inlet with simplified geometry. Since the focus of this work is to autotune the GPU kernel parameters on various types of GPU with different architectures, to maintain consistency and simplicity throughout the experimentation, the problem size is fixed at 3328$\times$1024 cells. To prevent the display of a congestion of refined grid lines, a coarse grid level is depicted for the 2D inlet flow in Fig.~\ref{inlet-grid}. The grid can be generated by solving 2D elliptic grid generation equations with Dirichlet boundary conditions. The boundary specifications for this grid are shown in Table~\ref{boundary_conditions}. The inflow Mach number is 4.0, with corresponding pressure and temperature values of 12270 Pa and 217 K, respectively.

\begin{table}[ht]
\centering
\caption{Boundary Conditions}
\label{boundary_conditions}
\begin{tabular}{c p{8cm}}
\toprule
\textbf{Boundary} & \textbf{Description} \\
\toprule
$\xi_{\text{min}}$ & $y = 0$ m for $-0.05$ m $\leq x \leq 0$ m \\
\hline
$\xi_{\text{max}}$ & $x = 1.5$ m for $0.2$ m $\leq y \leq 0.6$ m \\
\hline
$\eta_{\text{min}}$ & $y = x \tan(\theta)$ for $0 \leq x \leq 0.35$ m, where $\theta = 30$ deg. \\
 & $y = 0.2$ m for $0.35$ m $< x \leq 1.5$ m \\
\hline
$\eta_{\text{max}}$ & $x = (-0.05 + 3y^2 - 7.5y^4 + 9.5y^6)$ m for $0$ m $\leq y \leq 0.6$ m, $y = 0.6$ m for $0.50$ m $< x \leq 1.5$ m \\
\toprule
\end{tabular}
\end{table}

\begin{figure}[ht]
	\centering
	\includegraphics[width=.6\textwidth,trim=7 7 7 7,clip]{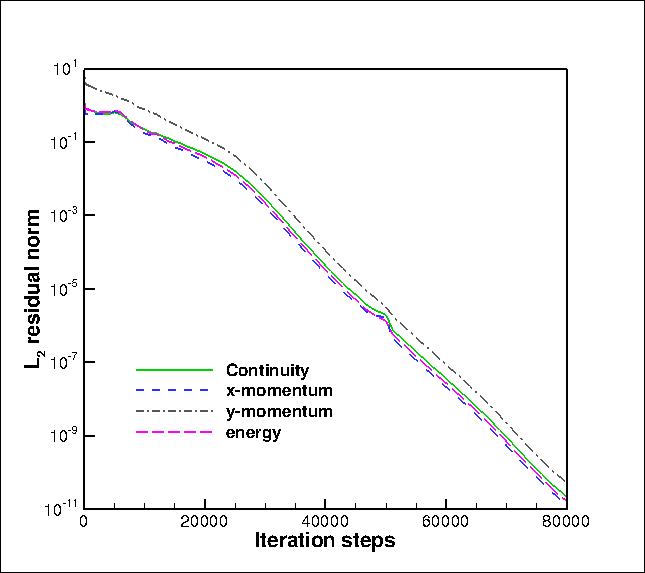}
	\caption{The relative iterative residual history for the inlet case}
	\label{inlet-resid}
\end{figure}

\begin{figure}[ht]
	\centering 
	\subfigure[Mach number contour and streamlines]{ 
		\label{inlet_Ma}
		\includegraphics[width=.45\textwidth,trim=7 7 7 7,clip]{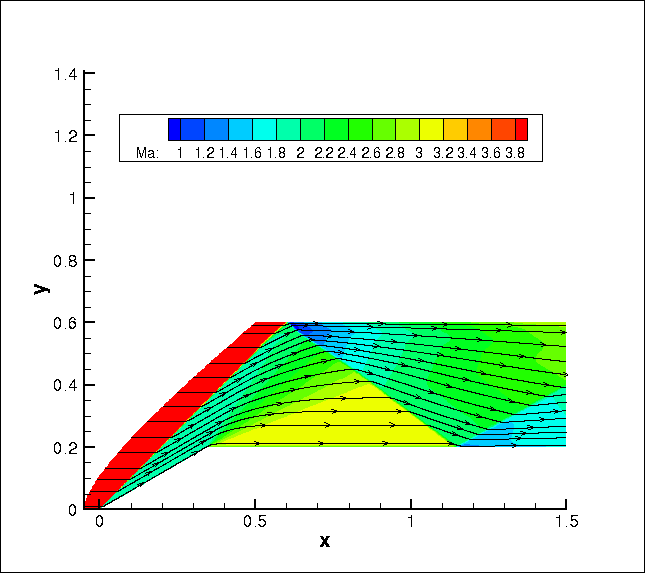} 
	} 
	\subfigure[Density contour]{ 
		\label{inlet_rho_euler}
		\includegraphics[width=.45\textwidth,trim=7 7 7 7,clip]{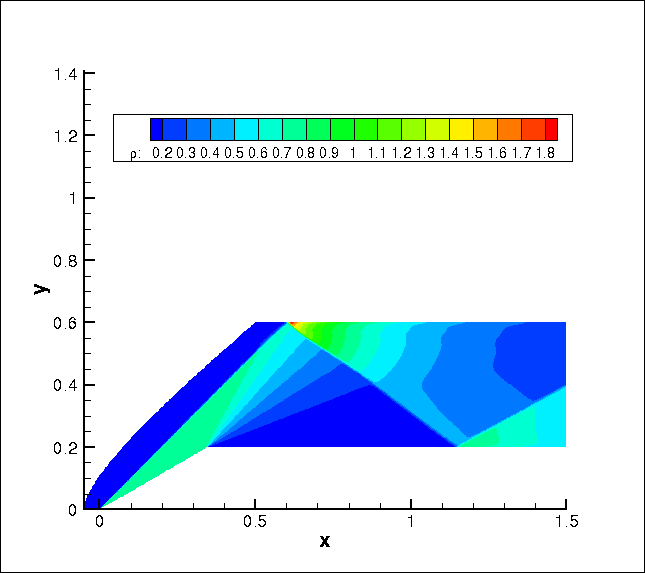} 
	} 
	\caption{2D supersonic inlet flow solution} 
	\label{Inlet-Mach-rho}
\end{figure}

Fig.~\ref{inlet-resid} displays the history of the relative residual $L_2$ norm using an explicit Runge-Kutta time marching method. The plot demonstrates that the iterative errors progressively decrease until convergence is achieved. At convergence, the iterative errors for all primitive variables have been sufficiently minimized, indicating the effectiveness of the CFD solver in accurately simulating the supersonic inlet problem. The parallel solution, utilizing 16 P100 GPUs, is illustrated in Fig.~\ref{Inlet-Mach-rho}. This test case serves as a representative example to evaluate and showcase the performance improvements achieved through our proposed autotuning approach.

\subsection{Hyperparameters}

In this work, we utilize the Scikit-learn~\cite{pedregosa2011scikit} framework for our machine learning autotuning tasks. Scikit-learn offers various useful functionalities, including the metric $R^2$, which is employed to evaluate the trained model's performance. The $R^2$ metric is defined in Eq.\ref{R_square}, and a value closer to 1 indicates a more accurate model. For optimizing the artificial neural network, the adaptive moment estimation (Adam) optimization algorithm is adopted. To determine the optimal configuration, a series of tests are conducted, leading to the selection of specific parameters for the neural network, as outlined in Table~\ref{para-setting}. The total number of layers, including hidden layers, along with the number of neurons in each layer, is an important aspect of the model design. The selection of these parameters are chosen based on a blend of heuristic principles guided by the official Scikit-learn tutorial and performance-driven tuning. Some hyperparameters are set by default, including the $L_2$ regularization coefficient $\alpha$ and the advam solver, while others undergo iterative adjustments to attain enhanced model performance, including the moment exponential dacay factors $\beta_1$ and $\beta_2$, initial learning rate, tolerance and numerical stability factor. Each layer's neuron count and the overall layering schema are tailored to optimize both training and inference phases. The selected hyperparameter values demonstrate optimal performance for the problem addressed in our study.

\begin{equation}
   \label{R_square}
   R^2 = 1 - \frac{\Sigma_{i=1}^N (y_{actual,i} - y_{pred,i})^2}{\Sigma_{i=1}^N (y_{actual,i} - y_{actual,mean})^2} 
\end{equation}

\begin{table}[ht]
  \caption{Hyperparameter for the artificial neural network}
  \label{para-setting}
  \centering
	\begin{tabular}{cc}
		\toprule
		Parameters                                          &
		Values   \\
		\toprule
		$L_2$ regularization coefficient, $\alpha$             & 0.0001     \\
		The 1st moment exponential decay factor, $\beta_1$  & 0.95       \\
		The 2nd moment exponential decay factor, $\beta_2$  & 0.90       \\
		Activation function                                 & Relu       \\
            Solver                                              & adam       \\
		Learning rate                                       &
		Adaptive   \\
		Initial learning rate                               & 0.0009     \\ 
		Maximum epoch number                                & 200        \\
		Batch size                                          & 200        \\		
		Tolerance                                           & $10^{-6}$  \\
		Numerical stability factor                          & $10^{-9}$  \\
		\bottomrule
	\end{tabular}
\end{table}

\subsection{Tuning Parameters}

To enhance the performance of SENSEI, our primary focus is on optimizing seven critical kernels that significantly contribute to the overall execution time. These kernels involve essential computations, such as limiter calculations in the $x$ and $y$ directions (xi limiter, eta limiter), flux calculations in the $x$ and $y$ directions (xi flux, eta flux), source term calculation, right-hand side (RHS) calculation, and solution updates. Each of these kernels is characterized by two tuning parameters: gang size and vector length, resulting in a total of 14 tuning parameters, as summarized in Table~\ref{tuning-parameters}. Specifically, each gang parameter in Table~\ref{tuning-parameters} lists 10 possible values, and each vector parameter lists 12 possible values.

The comprehensive search space for tuning these parameters is vast, amounting to $10^7 \times 12^7 = 3.58 \times 10^{14}$ possible combinations. This calculation arises from multiplying the available 10 values for each of the 7 gang parameters and 12 values for each of the 7 vector parameters, as detailed in Table~\ref{tuning-parameters}. However, owing to computational constraints, we are limited to utilizing only 7500 samples for training in this study.

Despite the restricted number of samples, our objective is to leverage the capabilities of machine learning to effectively explore and exploit this expansive search space. Through this approach, we aim to achieve improved performance for SENSEI by optimizing the selected kernels for the given tuning parameters. This strategy allows us to navigate the complexities of the tuning space efficiently, identifying configurations that lead to enhanced execution times and contributing to the overall optimization of the SENSEI application.

\begin{table}[ht]
  \caption{Tuning Parameters}
  \label{tuning-parameters}
  \centering
	\begin{tabular}{ll}
		\toprule
		Kernels  &  Range \\
		\toprule
		xi limiter gang   &  100, 200, ..., 1000\\
		xi limiter vector  &  32, 64, ..., 384\\
		eta limiter gang  &  100, 200, ..., 1000\\
		eta limiter vector  &  32, 64, ..., 384\\
		xi flux gang  &  100, 200, ..., 1000\\
		xi flux vector  &  32, 64, ..., 384\\
		eta flux gang  &  100, 200, ..., 1000\\
		eta flux vector  &  32, 64, ..., 384\\
		source term gang  &  100, 200, ..., 1000\\
		source term vector  &  32, 64, ..., 384\\
		right hand side gang  &  100, 200, ..., 1000\\
		right hand side vector  &  32, 64, ..., 384\\
		update solution gang  &  100, 200, ..., 1000\\
		update solution vector  &  32, 64, ..., 384\\
		\bottomrule
	\end{tabular}
\end{table}

\subsection{Feature Centering and Scaling}

In machine learning applications, it is common practice to apply feature centering and scaling, also known as mean removal and variance scaling. This technique offers several advantages, one of which is ensuring that different features are equally important by scaling them to the same range. This is crucial because it allows weights associated with different features to have a comparable magnitude. Failing to perform centering and scaling can lead to poor training performance. Scikit-learn provides various scaling methods, such as StandardScaler, MinMaxScaler, MaxAbsScaler, and more. In this work, we employ the StandardScaler, which is one of the simplest and most straightforward scalers. Although the StandardScaler assumes a Gaussian distribution for the data, it can still be beneficial even when the data distribution is unknown. By applying standard scaling, we aim to enhance the accuracy of our machine learning model, thereby improving the autotuning process in our study.

\subsection{GPU Model}

In this study, we utilize three distinct GPU accelerators, each offering varying speeds and capabilities to address a diverse range of computational workloads. A comprehensive comparison of these GPUs is presented in Table~\ref{accelerator}, showcasing key specifications such as peak floating-point precision performance, peak bandwidth, and memory size. Noteworthy is the deliberate inclusion of GPUs from different generations, spanning from the aging yet relevant C2075 GPU to the more recent and potent V100 GPUs. This strategic selection aims to ensure the efficacy of the machine learning autotuning approach across multiple GPU architectures.

These GPUs are built on different microarchitectures, reflecting the evolution of NVIDIA's GPU technology: the C2075 is rooted in the "Fermi" microarchitecture, the P100 is based on "Pascal," and the V100 leverages the advanced "Volta" microarchitecture.

The metric of double-precision performance, as detailed in Table~\ref{accelerator}, plays a pivotal role as an input feature for the artificial neural network, particularly in the combined training mode. This unique training approach involves using data from all three GPU types collectively. By doing so, the neural network gains the ability to learn and encapsulate the distinct performance characteristics of each GPU model, ranging from the legacy C2075 to the cutting-edge V100 GPUs. The incorporation of peak floating-point precision performance as a feature equips the neural network to make accurate runtime predictions across different GPU platforms. This adaptability is crucial for the autotuning approach to be effective across various GPU generations, thereby enhancing its applicability in diverse computing environments.

\begin{table}[ht]
  \caption{GPU specification}
  \label{accelerator}
  \centering
	\begin{tabular}{llll}
		\toprule
		GPU Model & C2075 & P100 & V100 \\
		\toprule
		Double precision performance, GFLOPS & 513 & 4700 & 7500 \\
        Peak bandwidth, GB/s & 144 & 720 & 900 \\
        Memory size, GB & 3 & 16 & 16 \\
		\bottomrule
	\end{tabular}
\end{table}

To account for the varying compute capabilities of different GPUs, the execution time of the program solver is adjusted to achieve a similar range across the GPUs. For instance, the inlet case is run for 1 iteration step on the C2075 GPU, 5 iteration steps on the P100 GPU, and 15 iteration steps on the V100 GPU. By adjusting the number of iteration steps, the solver time for all data samples is constrained to fall within the range of [0.8 s, 2.0 s]. It is important to note that the specified time range includes the execution time of the solver only and excludes any additional time spent on code launching, initial data movement, and final data movement. These additional time components are not considered since their impact can be compensated for by running more iteration steps. The entire data collection process involves running the program on each GPU platform for a total of 10,000 iterations. This process takes approximately 24 hours to complete on each GPU platform.

\section{Results}

\subsection{Training on Single Platform}

In this study, the initial training phase focuses on running SENSEI on individual GPUs, namely the C2075, V100, and P100 GPUs. The rationale behind this approach is to evaluate the practicality of employing artificial neural network techniques for auto-tuning a complex application, moving beyond simple kernels or benchmarks.

\begin{figure}[ht]
  \centering
  \subfigure[Training]{
  \includegraphics[width=.45\linewidth,  trim = 70 7 70 7, clip]{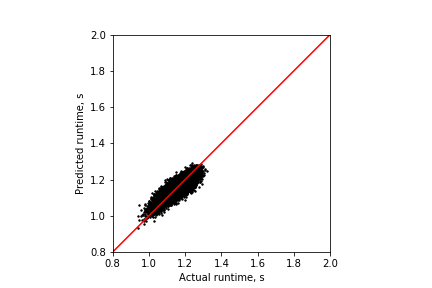}
  }
  \subfigure[Testing]{
  \includegraphics[width=.45\linewidth,  trim = 70 7 70 7, clip]{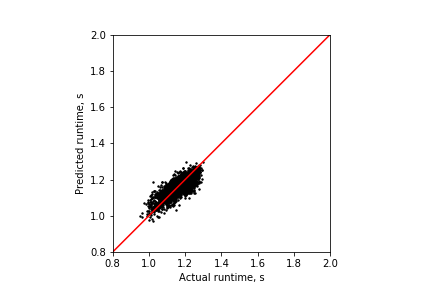}
  }
  \caption{Training and testing on C2075 GPU.}  
  \label{C2075_results}
\end{figure}

\begin{figure}[ht]
  \centering
  \subfigure[Training]{
  \includegraphics[width=.45\linewidth,  trim = 70 7 70 7, clip]{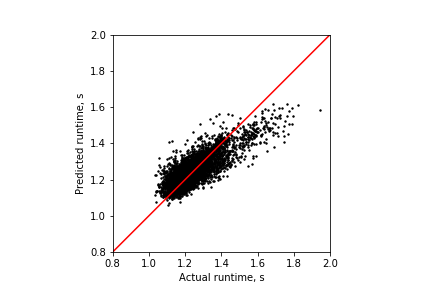}
  }
  \subfigure[Testing]{
  \includegraphics[width=.45\linewidth,  trim = 70 7 70 7, clip]{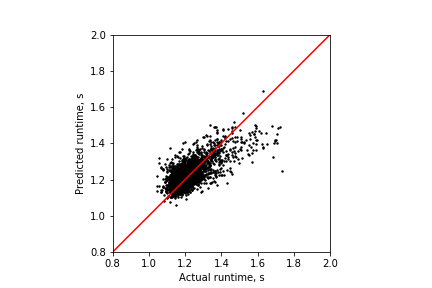}
  }
  \caption{Training and testing on V100 GPU.}  
  \label{V100_results}
\end{figure}

\begin{figure}[ht]
  \centering
  \subfigure[Training]{
  \includegraphics[width=.45\linewidth,  trim = 70 7 70 7, clip]{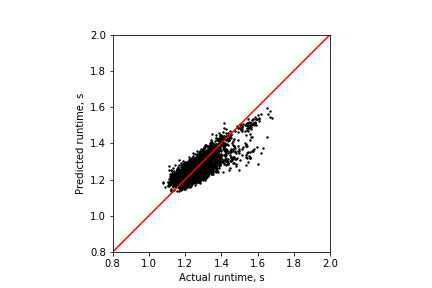}
  }
  \subfigure[Testing]{
  \includegraphics[width=.45\linewidth,  trim = 70 7 70 7, clip]{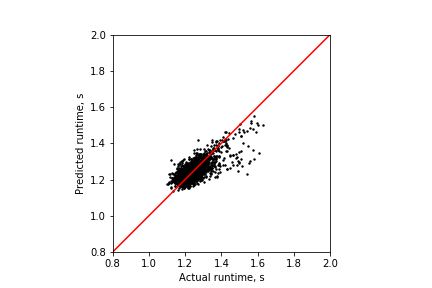}
  }
  \caption{Training and testing on P100 GPU.}  
  \label{P100_results}
\end{figure}

Fig.~\ref{C2075_results}, Fig.~\ref{V100_results}, and Fig.~\ref{P100_results} present the training and testing outcomes for each specific GPU. Each GPU dataset comprises 10,000 samples, with 7,500 allocated for training and the remaining 2,500 reserved for testing. The horizontal axis denotes the actual runtime, while the vertical axis represents the predicted runtime. The red line with a slope of 1 signifies perfect prediction for all samples. Notably, the neural network performs admirably across all tested GPUs, particularly excelling for slower GPUs where data points cluster closely around the perfect prediction line. Additionally, the model consistently identifies configurations with the lowest runtime, aligning closely with the actual best configurations. The focus here is on correctly identifying configurations with shorter actual runtimes, and while instances of large prediction errors exist for configurations with longer runtimes, the primary emphasis remains on achieving accurate predictions for shorter runtimes.

\subsection{Combined Training}

Moving to the subsequent phase, all data from diverse GPUs are amalgamated and subjected to combined training. The GPU mdel is incorporated as a feature to account for variations in computing capabilities among different GPUs. This unified training approach empowers the model to predict performance across a spectrum of GPUs, albeit without considering variations in problems, problem sizes, domain decompositions, or numerical schemes.

\begin{figure}[ht]
  \centering
  \includegraphics[width=.6\linewidth,  trim = 7 7 7 7, clip]{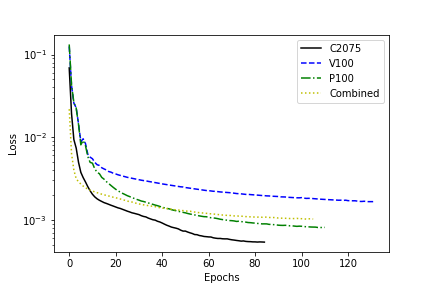}
  \caption{Loss history.}
  \label{loss-history}
\end{figure}

Fig.~\ref{loss-history} illustrates a comparison of the loss history between independent and combined training. The Mean Squared Error (MSE) loss function is employed. Notably, combined training exhibits a faster initial drop in loss, potentially attributed to improved centering and scaling facilitated by a larger dataset. The combined training utilizes a total of 22,500 samples, with each independent GPU contributing 7,500 samples, fostering enhanced stability and accuracy in the training process.

\begin{figure}[ht]
  \centering
  \includegraphics[width=.6\linewidth,  trim = 7 7 7 7, clip]{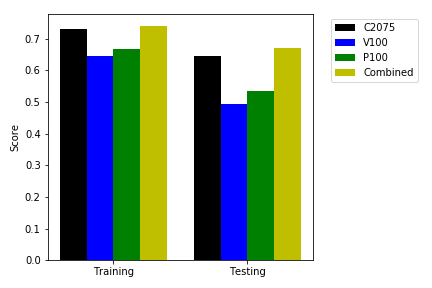}
  \caption{$R^2$ score.}
  \label{Score}
\end{figure}

Fig.~\ref{Score} further demonstrates the $R^2$ score comparison between independent and combined training. An intriguing observation is that a slower GPU exhibits a higher $R^2$ score than a faster GPU. However, upon consolidating all GPU data, both training and testing scores are significantly improved.

In conclusion, the findings highlight that combining data from various GPUs in the training process enhances prediction accuracy and performance assessment, surpassing the individual capabilities of each GPU.
    
\begin{figure}[htb]
  \centering
  \subfigure[Training]{
  \includegraphics[width=.45\linewidth,  trim = 70 7 70 7, clip]{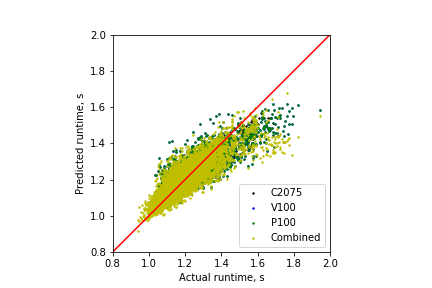}
  }
  \subfigure[Testing]{
  \includegraphics[width=.45\linewidth,  trim = 70 7 70 7, clip]{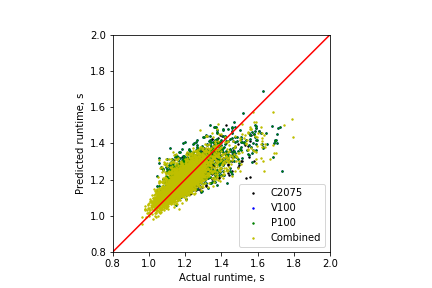}
  }
  \caption{Training and testing on combined dataset.}  
  \label{Combined_results}
\end{figure}

Fig.~\ref{Combined_results} underscores the success of the combined training and testing results for the 2D inlet case, showcasing the model's ability to accurately predict performance across different GPU types. This trained model stands as a valuable tool for estimating the performance of a diverse range of GPUs, offering insights into their expected runtime or efficiency for high-performance computing CFD simulations.

\section{Conclusions}

In this work, an artificial neural network approach is employed to automatically optimize fourteen GPU kernel scheduling parameters in a research CFD code accelerated by GPUs. Both independent training and combined training approaches are explored to assess the effectiveness of machine learning techniques in auto-tuning the code. Although this work focuses on a single problem with a fixed size, data from different GPU models are leveraged to predict the performance on a range of GPUs. This demonstrates the potential of utilizing machine learning to improve code optimization and performance tuning, even in scenarios with varying GPU architectures. 

\section*{Broader Impact}

Indeed, the findings of this work can have broader implications for general high performance scientific computing programs that require tuning and have a large search space. By leveraging machine learning techniques, these programs can potentially benefit from automated optimization and improved performance.

Furthermore, the development of compilers can be enhanced by incorporating similar machine learning-based toolkits on the backend. This would alleviate the burden on users to manually tune numerous parameters, which often necessitates expertise and is time-consuming. Such intelligent compiler systems could greatly simplify the optimization process and make it more accessible to a wider range of users.

To further enhance the practicality and significance of this research, future studies should consider multiple factors, including different types of problems, problem sizes, and numerical schemes. By taking these variables into account, the findings can be generalized to a broader range of scenarios and provide more comprehensive insights into code optimization and performance tuning in future works.

Additionally, extending this work to more GPUs would be valuable. Understanding how machine learning techniques perform when applied to a large number of GPU models in a heterogeneous computing environment would open up new possibilities for optimizing and scaling applications across diverse hardware configurations.

Exploring alternative methods, such as reinforcement learning, could also be worthwhile. Reinforcement learning algorithms have the potential to autonomously discover optimal tuning strategies by iteratively interacting with the system and learning from the feedback received.

By considering these suggestions and expanding the scope of the research, this work can become more meaningful, practical, and relevant to a wider range of applications and computing environments in future works.


\medskip

\small

\bibliographystyle{unsrtnat}
\bibliography{mybib}

\begin{thebibliography}{26}
\providecommand{\natexlab}[1]{#1}
\providecommand{\url}[1]{\texttt{#1}}
\expandafter\ifx\csname urlstyle\endcsname\relax
  \providecommand{\doi}[1]{doi: #1}\else
  \providecommand{\doi}{doi: \begingroup \urlstyle{rm}\Url}\fi

\bibitem[Hwu(2011)]{hwu2011gpu}
Wen-Mei~W Hwu.
\newblock \emph{{GPU Computing Gems Emerald Edition}}.
\newblock Elsevier, 2011.

\bibitem[Li et~al.(2019)Li, Song, Chen, Li, Liu, Tallent, and
  Barker]{li2019evaluating}
Ang Li, Shuaiwen~Leon Song, Jieyang Chen, Jiajia Li, Xu~Liu, Nathan~R Tallent,
  and Kevin~J Barker.
\newblock {Evaluating Modern GPU Interconnect: PCIe, NVLink, NV-Sli, NVSwitch
  and GPUDirect}.
\newblock \emph{IEEE Transactions on Parallel and Distributed Systems},
  31\penalty0 (1):\penalty0 94--110, 2019.

\bibitem[NVIDIA(2019)]{CUDA}
NVIDIA.
\newblock {CUDA C++ Programming Guide}, 2019.
\newblock URL \url{https://docs.nvidia.com/pdf/CUDA_C_Programming_Guide.pdf}.
\newblock (last accessed on 07/24/20).

\bibitem[{Khronos OpenCL Working Group}(2012)]{OpenCL}
{Khronos OpenCL Working Group}.
\newblock {The OpenCL Specification}, 2012.
\newblock URL
  \url{https://www.khronos.org/registry/OpenCL/specs/opencl-1.2.pdf}.
\newblock (last accessed on 07/24/20).

\bibitem[Ope(2015)]{OpenACC}
{OpenACC Programming and Best Practices Guide}, 2015.
\newblock URL
  \url{https://www.openacc.org/sites/default/files/inline-files/OpenACC_Programming_Guide_0.pdf}.
\newblock (last accessed on 07/24/20).

\bibitem[Mishra et~al.(2017)Mishra, Li, Kong, Finkel, and
  Chapman]{mishra2017benchmarking}
Alok Mishra, Lingda Li, Martin Kong, Hal Finkel, and Barbara Chapman.
\newblock {Benchmarking and Evaluating Unified Memory for OpenMP GPU
  Offloading}.
\newblock In \emph{Proceedings of the Fourth Workshop on the LLVM Compiler
  Infrastructure in HPC}, pages 1--10, 2017.

\bibitem[Sultana et~al.(2016)Sultana, Calvert, Overbey, and
  Arnold]{sultana2016openacc}
Nawrin Sultana, Alexander Calvert, Jeffrey~L Overbey, and Galen Arnold.
\newblock {From OpenACC to OpenMP 4: toward Automatic Translation}.
\newblock In \emph{Proceedings of the XSEDE16 Conference on Diversity, Big
  Data, and Science at Scale}, pages 1--8, 2016.

\bibitem[Chorin(1997)]{chorin1997numerical}
Alexandre~Joel Chorin.
\newblock A numerical method for solving incompressible viscous flow problems.
\newblock \emph{Journal of computational physics}, 135\penalty0 (2):\penalty0
  118--125, 1997.

\bibitem[Xue and Roy(2021)]{xue2021multi}
Weicheng Xue and Christoper~J Roy.
\newblock Multi-gpu performance optimization of a computational fluid dynamics
  code using openacc.
\newblock \emph{Concurrency and Computation: Practice and Experience},
  33\penalty0 (5):\penalty0 e6036, 2021.

\bibitem[Pickering et~al.(2015)Pickering, Jackson, Scogland, Feng, and
  Roy]{pickering2015directive}
Brent~P Pickering, Charles~W Jackson, Thomas~RW Scogland, Wu-Chun Feng, and
  Christopher~J Roy.
\newblock {Directive-based GPU Programming for Computational Fluid Dynamics}.
\newblock \emph{Computers \& Fluids}, 114:\penalty0 242--253, 2015.

\bibitem[Jia et~al.(2013)Jia, Shaw, and Martonosi]{jia2013starchart}
Wenhao Jia, Kelly~A Shaw, and Margaret Martonosi.
\newblock {Starchart: Hardware and Software Optimization using Recursive
  Partitioning Regression Trees}.
\newblock In \emph{Proceedings of the 22nd international conference on Parallel
  architectures and compilation techniques}, pages 257--267. IEEE, 2013.

\bibitem[Collins et~al.(2013)Collins, Fensch, Leather, and
  Cole]{collins2013masif}
Alexander Collins, Christian Fensch, Hugh Leather, and Murray Cole.
\newblock {MaSiF: Machine Learning Guided Auto-tuning of Parallel Skeletons}.
\newblock In \emph{20th Annual International Conference on High Performance
  Computing}, pages 186--195. IEEE, 2013.

\bibitem[Falch and Elster(2017)]{falch2017machine}
Thomas~L Falch and Anne~C Elster.
\newblock {Machine Learning-based Auto-tuning for Enhanced Performance
  Portability of OpenCL Applications}.
\newblock \emph{Concurrency and Computation: Practice and Experience},
  29\penalty0 (8):\penalty0 e4029, 2017.

\bibitem[Cui and Feng(2020)]{cui2020iterml}
Xuewen Cui and Wu-chun Feng.
\newblock {IterML: Iterative Machine Learning for Intelligent Parameter Pruning
  and Tuning in Graphics Processing Units}.
\newblock \emph{Journal of Signal Processing Systems}, pages 1--13, 2020.

\bibitem[Chen et~al.(2018)Chen, Moreau, Jiang, Zheng, Yan, Cowan, Shen, Wang,
  Hu, Ceze, et~al.]{chen2018tvm}
Tianqi Chen, Thierry Moreau, Ziheng Jiang, Lianmin Zheng, Eddie Yan, Meghan
  Cowan, Haichen Shen, Leyuan Wang, Yuwei Hu, Luis Ceze, et~al.
\newblock {TVM: An Automated End-to-end Optimizing Compiler for Deep Learning}.
\newblock \emph{arXiv preprint arXiv:1802.04799}, 2018.

\bibitem[Schoonhoven et~al.(2022)Schoonhoven, van Werkhoven, and
  Batenburg]{schoonhoven2022benchmarking}
Richard Schoonhoven, Ben van Werkhoven, and K~Joost Batenburg.
\newblock {Benchmarking Optimization Algorithms for Auto-tuning GPU Kernels}.
\newblock \emph{IEEE Transactions on Evolutionary Computation}, 2022.

\bibitem[Oberkampf and Roy(2010)]{oberkampf2010verification}
William~L Oberkampf and Christopher~J Roy.
\newblock \emph{Verification and validation in scientific computing}.
\newblock Cambridge University Press, 2010.

\bibitem[Allmaras and Johnson(2012)]{allmaras2012modifications}
Steven~R Allmaras and Forrester~T Johnson.
\newblock {Modifications and Clarifications for the Implementation of the
  Spalart-Allmaras Turbulence Model}.
\newblock In \emph{Seventh international conference on computational fluid
  dynamics (ICCFD7)}, pages 1--11, 2012.

\bibitem[Menter(1994)]{menter1994two}
Florian~R Menter.
\newblock {Two-equation Eddy-viscosity Turbulence Models for Engineering
  Applications}.
\newblock \emph{AIAA journal}, 32\penalty0 (8):\penalty0 1598--1605, 1994.

\bibitem[Menter et~al.(2003)Menter, Kuntz, and Langtry]{menter2003ten}
Florian~R Menter, Martin Kuntz, and Robin Langtry.
\newblock {Ten Years of Industrial Experience with the SST Turbulence Model}.
\newblock \emph{Turbulence, heat and mass transfer}, 4\penalty0 (1):\penalty0
  625--632, 2003.

\bibitem[Wang et~al.(2020)Wang, Xue, and Roy]{wang2020error}
Hongyu Wang, Weicheng Xue, and Christopher~J Roy.
\newblock {Error Transport Equation Implementation in the SENSEI CFD Code}.
\newblock In \emph{AIAA Scitech 2020 Forum}, page 1047, 2020.

\bibitem[Wang et~al.(2022)Wang, Xue, and Roy]{wang2022iterated}
Hongyu Wang, Weicheng Xue, and Christopher~J Roy.
\newblock {Iterated Discretization Error Transport Equations for Laminar and
  Turbulent Flows}.
\newblock \emph{International Journal for Numerical Methods in Fluids},
  94\penalty0 (6):\penalty0 536--556, 2022.

\bibitem[Xue et~al.(2021)Xue, Jackson, and Roy]{xue2021improved}
Weicheng Xue, Charles~W Jackson, and Christoper~J Roy.
\newblock {An Improved Framework of GPU Computing for CFD Applications on
  Structured Grids using OpenACC}.
\newblock \emph{Journal of Parallel and Distributed Computing}, 156:\penalty0
  64--85, 2021.

\bibitem[Xue et~al.(2023)Xue, Wang, and Roy]{xue2023cpu}
Weicheng Xue, Hongyu Wang, and Christopher~J Roy.
\newblock {CPU-GPU Heterogeneous Code Acceleration of a Finite Volume
  Computational Fluid Dynamics Solver}.
\newblock \emph{arXiv preprint arXiv:2305.18057}, 2023.

\bibitem[Feng(2018)]{feng2018deep}
Wu-Chun Feng.
\newblock {A Deep Learning Approach towards Auto Tuning CFD Codes}.
\newblock Technical report, Virginia Polytechnic Institute And State University
  Blacksburg United States, 2018.

\bibitem[Pedregosa et~al.(2011)Pedregosa, Varoquaux, Gramfort, Michel, Thirion,
  Grisel, Blondel, Prettenhofer, Weiss, Dubourg, et~al.]{pedregosa2011scikit}
Fabian Pedregosa, Ga{\"e}l Varoquaux, Alexandre Gramfort, Vincent Michel,
  Bertrand Thirion, Olivier Grisel, Mathieu Blondel, Peter Prettenhofer, Ron
  Weiss, Vincent Dubourg, et~al.
\newblock {Scikit-learn: Machine Learning in Python}.
\newblock \emph{the Journal of machine Learning research}, 12:\penalty0
  2825--2830, 2011.

\end{thebibliography}

\end{document}